# Standardized Evaluation of Automatic Methods for Perivascular Spaces Segmentation in MRI – MICCAI 2024 Challenge Results

Yilei Wu[*a], Yichi Zhang[*a], Zijian Dong[*a], Fang Ji[a], An Sen Tan[b], Gifford Tan[b], Sizhao Tang[b], Huijuan Chen[a], Zijiao Chen[a], Eric Kwun Kei Ng[a], Jose Bernal[e,f,p], Hang Min[g,q], Ying Xia[g], Ines Vati[g,r], Liz Cooper[g], Xiaoyu Hu[h], Yuchen Pei[h], Yutao Ma[h], Victor Nozais[i], Ami Tsuchida[j,k], Pierre-Yves Hervé[i], Philippe Boutinaud[i], Marc Joliot[j], Junghwa Kang[l], Wooseung Kim[l], Dayeon Bak[l], Rachika E. Hamadache[n], Valeriia Abramova[n], Xavier Lladó[n], Yuntao Zhu[o], Zhenyu Gong[d], Xin Chen[m], John McFadden[e], Pek Lan Khong[b], Roberto Duarte Coello[e], Hongwei Bran Li[c], Woon Puay Koh[s], Christopher Chen[t], Joanna M. Wardlaw[e], Maria del C. Valdés Hernández[†e], Juan Helen Zhou[†a,s,u,v]

[a] Centre for Sleep and Cognition (CSC) & Centre for Translational Magnetic Resonance Research (TMR), Yong Loo Lin School of Medicine, National University of Singapore, Singapore [b] National University Hospital, Singapore [c] Harvard Medical School, Boston, Massachusetts, USA [d] Department of Neurosurgery, Klinikum Rechts der Isar, Technical University of Munich, Germany [e] Centre for Clinical Brain Sciences, UK Dementia Research Centre, The University of Edinburgh, Edinburgh, UK [f] German Centre for Neurodegenerative Diseases (DZNE), Germany [g] Australian e-Health Research Centre, CSIRO Health and Biosecurity, Herston, 4029, Queensland, Australia [h] Central China Normal University, Wuhan, China [I] Fealinx, France [j] Groupe d'Imagerie Neurofonctionnelle (GIN), Institute of Neurodegenerative Diseases (IMN), UMR5293, CNRS, CEA, University of Bordeaux, Bordeaux, France [k] Bordeaux Population Health, INSERM, U1219, University of Bordeaux, Bordeaux, France [l] Department of Biomedical Engineering, Hankuk University

of Foreign Studies, Yongin, Korea [m] Longhua Hospital Shanghai University of Traditional Chinese Medicine, Shanghai, China [n] Research Institute of Computer Vision and Robotics (ViCOROB), Universitat de Girona, Catalonia, Spain [o] School of Mathematics, Nanjing University, Nanjing, China [p] Institute of Cognitive Neurology and Dementia Research (IKND), Otto von Guericke University Magdeburg, Magdeburg, Germany [q] South Western Clinical School, University of New South Wales, Sydney, Australia [r] School of Electrical Engineering and Robotics, Queensland University of Technology, Brisbane, Australia [s] Healthy Longevity Translational Research Program, Yong Loo Lin School of Medicine, National University of Singapore, Singapore [t] Department of Pharmacology, National University of Singapore, Singapore [u] Human Potential Translational Research Program and Department of Medicine, Yong Loo Lin School of Medicine, National University of Singapore, Singapore [v] Department of Electrical and Computer Engineering, National University of Singapore, Singapore

**Address correspondence to:**

J. Helen Zhou

CSC, TMR, HLTRP, HPTRP, ECE

National University of Singapore

Email: helen.zhou@nus.edu.sg

Maria Valdés Hernández

CCBS

The University of Edinburgh

Email: m.valdes-hernan@ed.ac.uk

---

[1]* Equal contribution

[2]† Corresponding author

## Abstract

Perivascular spaces (PVS), when abnormally enlarged and visible in magnetic resonance imaging (MRI) structural sequences, are important imaging markers of cerebral small vessel disease and potential indicators of neurodegenerative conditions. Despite their clinical significance, automatic enlarged PVS (EPVS) segmentation remains challenging due to their small size, variable morphology, similarity with other pathological features, and limited annotated datasets. This paper presents the EPVS Challenge organized at MICCAI 2024, which aims to advance the development of automated algorithms for EPVS segmentation across multi-site data. We provided a diverse dataset comprising 100 training, 50 validation, and 50 testing scans collected from multiple international sites (UK, Singapore, and China) with varying MRI protocols and demographics. All annotations followed the STRIVE protocol to ensure standardized ground truth and covered the full brain parenchyma. Seven teams completed the full challenge, implementing various deep learning approaches primarily based on U-Net architectures with innovations in multi-modal processing, ensemble strategies, and transformer-based components. Performance was evaluated using dice similarity coefficient, absolute volume difference, recall, and precision metrics. The winning method employed MedNeXt architecture with a dual 2D/3D strategy for handling varying slice thicknesses. The top solutions showed relatively good performance on test data from seen datasets, but significant degradation of performance was observed on the previously unseen Shanghai cohort, highlighting cross-site generalization challenges due to domain shift. This challenge establishes an important benchmark for EPVS segmentation methods and underscores the need for the continued development of robust algorithms that can generalize in diverse clinical settings.

# 1. Introduction

Perivascular spaces (PVS) in humans are passageways that surround cerebral blood vessels, through which CSF can flow (Yamamoto et al., 2024). They are thought to be part of the brain's glymphatic system facilitating interstitial fluid exchange and clearance of waste products (Bown et al., 2022; Brown et al., 2018; Ramaswamy et al., 2022; Wardlaw et al., 2020). Enlarged perivascular spaces (EPVS) are common in older adults and can be detected in MRI as hyperintense linear features in T2-weighted images, or as hypointense features in T1-weighted and T2-weighted fluid attenuated inversion recovery (T2-FLAIR) images (Barisano et al., 2022; Moses et al., 2023; Van Veluw et al., 2022). Previous studies (Doubal et al., 2010; Smith et al., 2019; Wardlaw et al., 2013) have linked EPVS burden with cerebral small vessel disease (CSVD), diabetes mellitus, and cognitive impairment. Emerging evidence has shown the correlation of PVS with cognitive decline and dementia (Chen et al., 2024, Menze et al., 2024, Zhang et al., 2024). A longitudinal study found that severe PVS in both the basal ganglia (BG) and centrum semiovale (CSO) regions predicted greater cognitive decline and increased risk of dementia, independent of other small vessel disease markers (Paradise et al., 2021). Complementing this, a cross-sectional study demonstrated that PVS are independently associated with poorer cognitive performance, particularly in executive function and processing speed, beyond the effects of other CSVD markers (Passiak et al., 2019). Studies suggest that EPVS in different brain regions exhibit different correlations with other disease manifestations, cementing the hypothesis that different brain regions might have different underlying small vessel arteriopathies to which EPVS may be functionally linked (Hurford et al., 2014).

Quantifying EPVS is crucial in clinical research studies, as measures of EPVS count, distribution, and morphology have been associated with various neurological conditions, including CSVD and sleep apnea (Waymont et al., 2024), supporting diagnosis, prognosis, and disease monitoring in clinical practice. Currently, the assessment of EPVS still largely relies on coarse-grained visual scoring systems (Laveskog et al., 2018; Patankar et al., 2005; Potter et al., 2015) based on the number of EPVS counted in specific clinically relevant regions. While visual rating scales are relatively easy to implement in clinical settings, they are inherently subjective, suffering from sub-optimal inter-rater reliability and may not accurately capture the regional heterogeneity of PVS throughout the brain. Although detailed EPVS segmentation can provide more comprehensive quantitative measures for clinical analysis, manual segmentation remains a time-consuming and observer-dependent procedure (Moses et al., 2023; Waymont et al., 2024; Dubost et al., 2019; Ballerini et al., 2018).

Automatic EPVS segmentation methods (Waymont et al., 2024) have been developed, including conventional (Ballerini et al., 2018; Zhang et al., 2017) and deep learning-based (Cai et al., 2024; Zhang et al., 2017) methods. Each proposed segmentation method has been evaluated against a different ground truth (different number of participants, different experts, different protocols, different nature (e.g., visual scores, visual counts, reference segmentations), using different evaluation criteria. Therefore, it is hard to compare various methods.

Another challenge that stands out is the lack of annotated public datasets. In 2021, the VALDO challenge (https://valdo.grand-challenge.org/) (Sudre et al., 2024) focused on the detection of CSVD neuroimaging markers and provided 40 MRI volumes with EPVS masks from two European population-based studies as the training set. However, only 12 of them were annotated for all brain slices while the remaining 28 were labeled only for critical slices. This limited dataset size and site variability pose significant challenges for developing robust and generalizable

EPVS segmentation methods, particularly for deep learning-based approaches that typically require large amounts of fully annotated training data.

The primary objective of our 2024 MICCAI EPVS challenge was to develop and evaluate automated segmentation algorithms for accurate detection and delineation of EPVS in brain MRI scans across multiple sites, ethnicities and disease conditions. Participants in the challenge were tasked with developing segmentation models that can generate binary EPVS masks from multi-modal MRI inputs including T1-weighted, T2-weighted, and T2-FLAIR sequences. To facilitate algorithm development and ensure robust evaluation, we provided a dataset comprising 100 fully annotated training scans, 50 validation scans, and 50 testing scans collected from three international sites from the UK, Singapore and China. The scans were acquired with varying MRI protocols, scanner types, and with different patient demographics. All EPVS annotations strictly followed the Standards for Reporting Vascular Changes on Neuroimaging (STRIVE) protocol (Wardlaw et al., 2013), ensuring standardized and reliable ground truth labels. By incorporating this multi-site, multi-scanner, and demographically diverse dataset, we aimed to motivate the development of robust algorithms that can generalize across different clinical and research settings, ultimately advancing the field towards achieving reliable automated EPVS quantification for real-world applications. This paper describes the design, results and lessons learned through the challenge according to the reporting guidelines detailed in Maier-Hein et al., 2020.

## 2. Methods

### 2.1. Challenge organization

The Enlarged Perivascular Spaces (EPVS) Segmentation Challenge was hosted as a satellite event at MICCAI 2024, organized through a collaboration between the National University of Singapore, the National University Hospital, the University of Edinburgh, and the Longhua Hospital Shanghai University of Traditional Chinese Medicine. The challenge design was peer reviewed by the MICCAI 2024 challenge organization committee and all the documentation was made publicly available from https://zenodo.org/records/10992174. To ensure fairness, organizing team members were ineligible for participation and prizes. Awards were distributed to winners of the individual ROI categories as well as the overall challenge winner, with the results publicly displayed on the challenge website. Teams that submitted valid entries were invited to nominate up to three members as co-authors for the challenge overview paper. Following the publication of this overview paper, the challenge platform will remain open for researchers to benchmark new methods against existing submissions. As shown in **Figure 1**, the EPVS challenge dataset consists of co-registered T1-weighted, T2-weighted, and T2-FLAIR scans as well as the EPVS manual masks in the ROI of CSO and BG for each subject. The dataset is accessible upon request following the terms of use. Besides, a synthetic dataset developed for this challenge is publicly available from the hosting GitHub repository (https://github.com/joseabernal/PVSDRO). The challenge website (https://www.synapse.org/Synapse:syn54100278/wiki/626542) provides comprehensive information about the competition, including evaluation criteria, submission guidelines, and ongoing developments.

As shown in **Table 1**, the challenge was organized in three phases: (1) a training phase from the moment the fully annotated real and synthetic datasets for training were made available; (2) a validation phase when we released the validation dataset to provide participants with individual performance feedback using a leaderboard; (3) the final evaluation stage on unseen test cases. Participants were asked to provide a docker container for their fully automated methods, and organizers would contact participants individually via email if technical issues occurred. Details of the algorithm submission procedure are specified in https://www.synapse.org/Synapse:syn54100278/wiki/628973 and https://www.synapse.org/Synapse:syn54100278/wiki/628975. Participating teams were also requested to provide a short technical note describing their solutions, which are also available from https://www.synapse.org/Synapse:syn64956514.

The code for evaluation metrics was made available prior to submission at https://github.com/hzlab/EPVS_challenge. Participating teams were encouraged to make their source code publicly available, and all participants agreed that docker containers could be made public from https://hub.docker.com/u/epvschallenge.

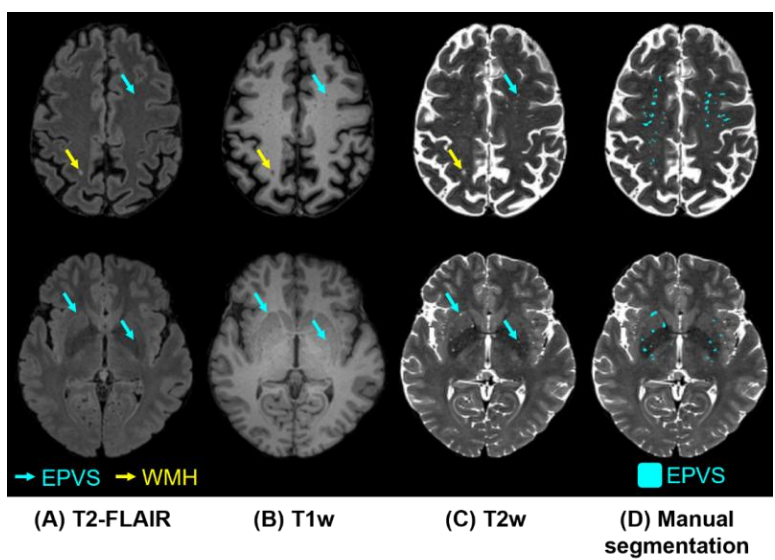

**Figure 1. Axial brain MRIs demonstrating EPVS at two anatomical levels at the CSO (top) and BG (bottom).** The imaging sequences include T2-FLAIR **(A)**,

T1-weighted **(B)**, and T2-weighted **(C)** MRIs. Manual EPVS annotations are shown as cyan masks overlaid on the T2-weighted image **(D)**, highlighting the distribution of EPVS in the CSO and BG regions. **Abbreviations:** EPVS – enlarged perivascular spaces; WMH – white matter hyperintensities; CSO – centrum semiovale; BG – basal ganglia.

| Phase | Start Date | End Date | Materials Released |
|---|---|---|---|
| Training phase | May 15, 2024 | June 14, 2024 | Training data (with binary masks) (100 cases) |
| Validation phase | June 15, 2024 | Sept 15, 2024 | Validation data (50 cases) |
| Test phase | Aug 15, 2024 | Sept 15, 2024 | None |
| Challenge Day | Oct 10, 2024 | Oct 10, 2024 | Final result |

**Table 1: Key dates and materials for different phases of the EPVS Challenge.**

## 2.2. Challenge dataset and acquisition

### 2.2.1. Challenge data cohorts

The EPVS challenge dataset comprised diverse MRI data collected from multiple international sites, including the UK (University of Edinburgh), Singapore (SG70, MACC), and China (Shanghai TCM cohort). This comprehensive collection represented a wide spectrum of demographic and clinical profiles, ranging from healthy elderly individuals with varying cognitive performance to patients with conditions such as sleep apnea, mild stroke, and dementia. For each participant, T1-weighted, T2-weighted, and T2-FLAIR MRI images were collected and provided. **Figure 2** illustrates the variability of the T2-weighted MRIs and EPVS annotations of the data from three sites. A brief introduction of each site is given below.

**University of Edinburgh** The Edinburgh dataset contributed data from four distinct cohorts: (1) The Lothian Birth Cohort 1936 Study included community-dwelling, cognitively normal individuals born in Edinburgh and The Lothians regions in 1936, with brain MRI scans acquired at mean age 72.6 years, and EPVS and white matter hyperintensity (WMH) burdens ranging from none to moderate (Wardlaw et al., 2011); (2) A sleep study cohort comprising participants with moderate to severe obstructive sleep apnea (OSA) but otherwise healthy, with mean age 50.4 years and generally abundant EPVS burden despite minimal vascular pathology (Clancy et al., 2021); (3) An Observational Study of Mild Stroke including patients with mild-to-moderate ischemic stroke (predominantly lacunar type), mean age 66 years, presenting abundant vascular pathology and moderate to severe EPVS burden (Clancy et al., 2021); and (4) The Stratifying Resilience and Depression Longitudinally (STRADL) study, a depression-focused investigation using data from Generation Scotland, featuring healthy community-dwelling adults aged 18+ with none to moderate vascular pathology and EPVS burden (Habota et al., 2021).

**SG70** The SG70 dataset was a community-based subset of the Singapore Chinese Health Study (SCHS) cohort (Lai et al., 2024), comprising city-dwelling healthy elderly individuals with varying levels of cognitive performance, with mean age of 72.8 years. Participants were recalled for brain imaging, contributing valuable data for studying pre-clinical markers of cognitive decline in an Asian population.

**MACC** The MACC dataset was a Singaporean cohort (Ji et al., 2023) selected from the Memory Ageing and Cognition Centre cohort recruited from the memory clinics of the National University Hospital and St. Luke's Hospital in Singapore (Van Veluw et al., 2015). This cohort represented various stages of cognitive decline, including participants spanning from no cognition impairment (NCI), mild cognitive impairment (MCI) and dementia, mean age of 65.8. This cohort provides a diverse range of cognitive profiles for research and analysis in the context of neurodegenerative conditions.

**Shanghai TCM** The Shanghai TCM dataset was a community-based cohort collected at Shanghai No. 6 People's Hospital and Longhua Hospital, Shanghai University of Traditional Chinese Medicine. It consisted of healthy elderly individuals at risk of dementia with cognitive performance at the mean level or slightly lower than expected for their age, with mean age of 74.1 years. Compared with the other three research-oriented sites, this site collected data under a clinical setting. Both hospitals have multiple scanners of different manufactures, and the scans followed clinical protocols with large slice spacing and small pixel spacing, allowing for an estimation of performance degradation due to domain shift.

The training data consisted of 100 images: 60 images from the University of Edinburgh which span across multiple cohorts and scanner configurations, and 20 images from two Singaporean cohorts each. The validation data included 30 images from the University of Edinburgh and 10 from two Singapore cohorts each. The test data consisted of 50 images, and each site of the University of Edinburgh, SG70 and MACC contributed 10 images, while all the 20 images from the Shanghai TCM site were withheld as test data to evaluate the generalizability on unseen datasets under a clinical setting. The split was designed to ensure adequate training while maintaining sufficient data for robust validation and challenging test evaluation across multiple sites. An overview of the imaging parameters of three sites is provided in **Table 2**. The distribution of EPVS features, such as total volume and cluster count, varied considerably across sites due to cohort and scanner diversity, as illustrated in **Figure 3**. The senior cohort and the high axial resolution lead to more detections of EPVS as large clusters on the Shanghai TCM site while EPVS appears as small blobs or thin lines on other three sites as shown in **Figure 2**, highlighting the diversity and challenge of the dataset.

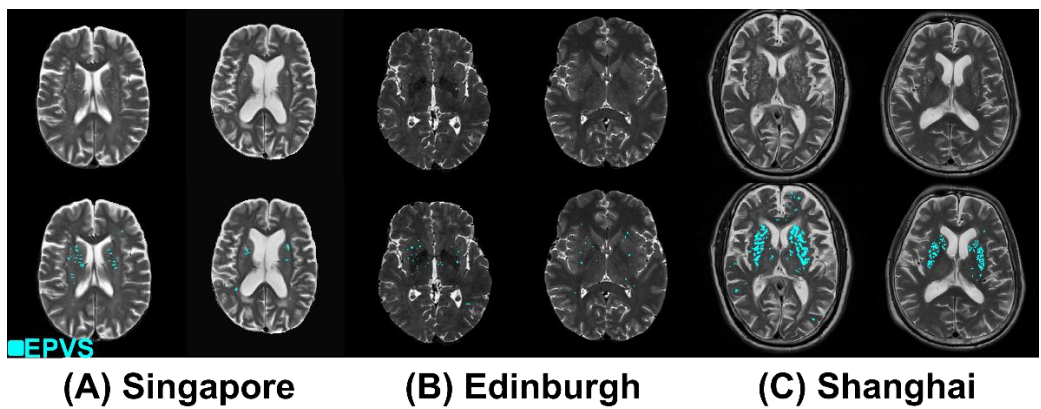

**Figure 2: Sample T2-weighted brain MRI images from three sites in the EPVS challenge. (A)** Singapore, **(B)** Edinburgh, and **(C)** Shanghai. Manual EPVS annotations are shown as cyan masks overlaid on the T2-weighted image at the bottom row, demonstrating the varied appearance and distribution of EPVS across different sites. **Abbreviations**: EPVS – enlarged perivascular spaces.

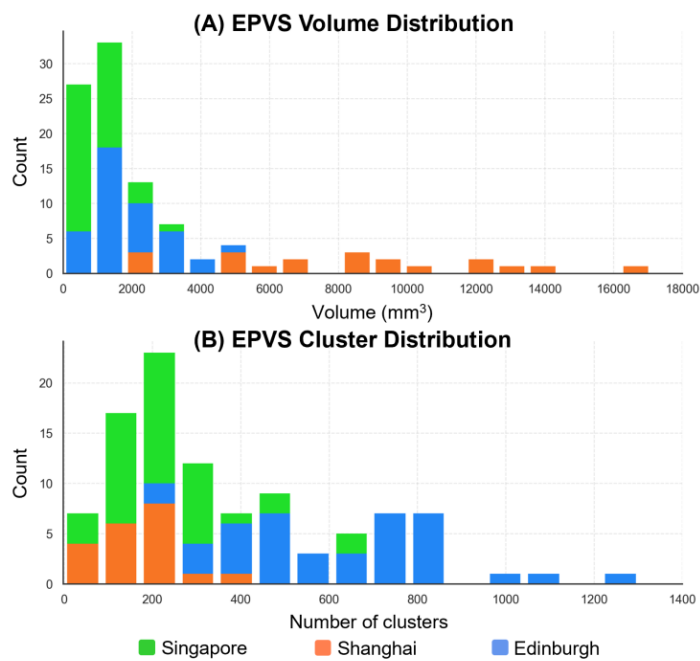

**Figure 3: Distribution of EPVS characteristics across three sites. (A)** EPVS volume distribution showing Singaporean samples (green) predominantly in the lower volume ranges (below 2000 mm³), Edinburgh samples (blue) spanning moderate volumes (500-4000 mm³), and Shanghai samples (orange) extending

into higher volume ranges (above 5000 mm³). **(B)** EPVS cluster (defined as a group of connected voxels that form a single continuous structure in 3D space) count distribution revealing Singaporean and Shanghai samples concentrated in lower cluster ranges (below 350), while Edinburgh samples demonstrate higher cluster counts (many between 500-1050). This pattern suggests the diversity of both the cohort and the scanners from three sites. **Abbreviations:** EPVS – enlarged perivascular spaces.

| Dataset | Scanner | Modality | Voxel Size (mm³) | Train | Val | Test |
|---|---|---|---|---|---|---|
| University of Edinburgh | 1.5T GE Signa Horizon | T1w | $1 \times 1 \times 1.3$ | 60 | 30 | 10 |
| | | T2w | $1 \times 1 \times 2$ | | | |
| | | T2-FLAIR | $1 \times 1 \times 4$ | | | |
| | 3T Siemens Prisma | T1w | $1 \times 1 \times 1$ | | | |
| | | T2w | $0.9 \times 0.9 \times 0.9$ | | | |
| | | T2-FLAIR | $1 \times 1 \times 1$ | | | |
| | 3T Philips Achieva | T1w | $1 \times 1 \times 1$ | | | |
| | | T2w | $0.5 \times 0.5 \times 0.5$ | | | |
| | | T2-FLAIR | $0.94 \times 0.94 \times 1$ | | | |
| | 3T Siemens Prisma | T1w | $1 \times 1 \times 1$ | | | |
| | | T2w | $0.5 \times 0.5 \times 0.5$ | | | |
| | | T2-FLAIR | $1 \times 1 \times 1$ | | | |
| NUS-SG70 | 3T Siemens Prisma | T1w | $1 \times 1 \times 1$ | 20 | 10 | 10 |
| | | T2w | $1 \times 1 \times 3$ | | | |
| | | T2-FLAIR | $1 \times 1 \times 1$ | | | |
| NUS-MACC | 3T Siemens Prisma | T1w | $1 \times 1 \times 1$ | 20 | 10 | 10 |
| | | T2w | $1 \times 1 \times 3$ | | | |
| | | T2-FLAIR | $1 \times 1 \times 3$ | | | |
| Shanghai TCM | 1.5T Siemens Aera | T1w | $0.45 \times 0.45 \times 7.8$ | 0 | 0 | 20 |
| | | T2w | $0.72 \times 0.72 \times 7.2$ | | | |
| | | T2-FLAIR | $0.9 \times 0.9 \times 7.8$ | | | |
| | 1.5T Siemens Avanto | T1w | $0.9 \times 0.9 \times 7.5$ | | | |
| | | T2w | $0.36 \times 0.36 \times 7.5$ | | | |
| | | T2-FLAIR | $0.45 \times 0.45 \times 7.5$ | | | |
| | 3T Siemens Skyra | T1w | $0.72 \times 0.72 \times 6$ | | | |
| | | T2w | $0.72 \times 0.72 \times 6$ | | | |
| | | T2-FLAIR | $0.9 \times 0.9 \times 6$ | | | |
| | 3T GE Signa Pioneer | T1w | $0.47 \times 0.47 \times 6.5$ | | | |
| | | T2w | $0.47 \times 0.47 \times 6.5$ | | | |
| | | T2-FLAIR | $0.47 \times 0.47 \times 6.5$ | | | |
| | 3T Philips Ingenia | T1w | $0.53 \times 0.53 \times 6$ | | | |
| | | T2w | $0.36 \times 0.36 \times 6$ | | | |
| | | T2-FLAIR | $0.6 \times 0.6 \times 6$ | | | |

**Table 2: Overview of the imaging parameters of three sites.** The numbers of samples in training, validation (Val) and testing stages are provided in the last three columns.

### 2.2.2. Data Preprocessing

All original images were converted from DICOM to NIFTI format using either dcm2niix (Li et al., 2016) or mriconvert (Fischl, 2012). For each individual, images of all three modalities were co-registered to the native space with the highest resolution (SG70/MACC, T1-weighted space; Edinburgh/Shanghai TCM, T2-weighted space) using FSL-FLIRT (Jenkinson et al. 2002), and skull stripping was performed using either freesurfer (Fischl, 2012) or BET2 (Smith, 2002). All intracranial volumes were manually edited for accuracy. For the Singapore (SG70, MACC) and Shanghai TCM datasets, we employed boundary-based co-registration (Greve and Fischl., 2009) to improve alignment accuracy. PVS regions of interest (ROI) and segmentation priors were generated using a cohort-specific tailored modification of the pipeline developed by Valdes Hernández et al., 2023, which is publicly available from https://doi.org/10.7488/ds/7486.

### 2.2.3. Dataset Ethics

**University of Edinburgh**: All primary studies that provided data for this challenge were approved by the corresponding ethics boards and have been conducted according to the Declaration of Helsinki. All individuals gave written consent for their data to be used for research purposes. The data provided and used in this challenge are double-anonymised brain image data. We provided brain-extracted T1-weighted, T2-weighted and T2-FLAIR volumes (i.e., 3D volumes of image data only inside the cranial vault), co-registered to the T2-weighted native space. In addition, we provided binary masks for the regions of interest BG and CSO and binary masks of the manually edited PVS segmentation. Study IDs were also removed and specific IDs for this challenge were given. Data were grouped by MRI protocol, not by primary study source. Hence, challenge participants and the rest of the challenge organizers will be blind to primary study characteristics.

**MACC**: The MACC study had the ethics approval from the NHG DOMAIN SPECIFIC REVIEW BOARD (DSRB) APPROVAL with a reference number of 2015/00406. The NHG DSRB operated in accordance with the ICH GCP and all applicable laws and regulations. Participants were recruited into the study from NUHS memory clinics and NUHS MACC research cohorts.

**SG70**: The SG70 study obtained ethics approval from the NUS-IRB REVIEW BOARD APPROVAL with a reference number of 2020/398 and the DSRB APPROVAL with a reference number of 2020/01465. In addition to ICH GCP, this human biomedical research was regulated by the Human Biomedical Research Act (HBRA), and researchers were required by law to comply with all relevant regulatory requirements of the HBRA. The study was conducted at NUS - Yong Loo Lin School of Medicine.

**Shanghai TCM**: The Shanghai TCM cohort study received ethics approval from the Ethics Committee of Longhua Hospital Shanghai University of Traditional Chinese Medicine with approval number xjc22022244. All participants provided written informed consent before enrollment in the study. The data collection process adhered to appropriate privacy protection measures and data security protocols.

**2.2.4. Synthetic data**

A comprehensive synthetic dataset of 450 samples with ground truth annotations was generated by adapting the digital reference object framework available at https://github.com/joseabernal/PVSDRO, as described in (Bernal et al., 2022). It incorporated synthetic brain image data with varying degrees and types of noise and artefacts, including all standard neuroimaging modalities (T1-weighted, T2-weighted and T2-FLAIR sequences) of synthetic scans with five different EPVS lengths and three different diameters. The synthetic data were based on brain templates of multiple ages (25, 35, 45, 55, 73, 76, and 92 years) sourced from https://datashare.ed.ac.uk/handle/10283/8793 and representing different degrees of brain atrophy and tissue integrity. The participants were encouraged to use this synthetic dataset to improve their models.

**2.3. Data Annotation**

Our EPVS annotation pipeline adhered to the STRIVE standards (Wardlaw et al., 2013) and focused on three key brain regions: hippocampus, BG, and CSO, with masks of regions of interest derived from standardized atlases (Dickie et al., 2016).

For the Edinburgh cohort, we utilized the method described in (Ballerini et al., 2018) to generate initial segmentation priors. After that, all segmentation priors underwent thorough manual editing by two different trained observers who worked independently and were blinded to each other's edits. The final manual segmentation masks were determined through consensus among these annotators, and a third experienced annotator resolved any discrepancies. This ensured that the computational approach merely facilitated the manual annotation process rather than replacing expert knowledge.

Similarly, for the Singaporean cohorts (SG70 and MACC), we employed a pretrained U-Net (Boutinaud et al., 2021) to generate initial segmentation priors. Like the Edinburgh dataset, experienced neuroradiologists and neuroimaging researchers reviewed and manually edited these priors. If consensus was not achieved, a third annotator reviewed the manual segmentation masks to ensure accurate EPVS identification according to established criteria.

The Shanghai TCM Hospital dataset followed a different approach, with complete manual annotation from scratch without automated priors. Experienced neuroradiologists manually identified and segmented EPVS in all participants.

**Sources of annotation errors.** To ensure consistency across sites despite the different initial approaches, we hosted several discussion sessions where representatives of each center collectively reviewed cases, allowing for mutual feedback and better alignment. Despite all the effort, the annotation of EPVS faced several significant challenges that contributed to potential errors. EPVS structures are inherently difficult to identify due to their small size (typically only 1-3mm in diameter) and their similar appearance to surrounding tissue, making annotators prone to mislabeling and confusing with other small hypointensities such as lacunes or artifacts. This challenge was compounded by variations in the software used to generate segmentation priors, with methods like Ballerini's approach (Ballerini et al., 2018) and U-Net architectures (Boutinaud et al., 2021) each introducing their own biases in detection sensitivity and boundary definition. Further complicating accuracy are the limitations of manual annotation tools, including issues with brush size precision, variations in EPVS visibility across different viewing planes, limitations of single versus multiplanar viewing options, and interface sensitivity problems that affect the precise delineation of these minuscule structures. Together, these factors create considerable potential for inconsistency in EPVS annotation despite rigorous quality control measures.

## 2.4. Evaluation method

All methods were evaluated using four metrics: (1) dice similarity coefficient (DSC), (2) absolute volume difference (AVD), (3) sensitivity for detecting individual EPVS (Recall), and (4) positive predictive value for detecting individual EPVS (Precision). For recall and precision, individual EPVS were defined as 3D-connected components within an MRI volume. The exact implementation of each metric was put online beforehand and could be used by participants for self-evaluation during development. In extreme cases of false positives, huge volume difference between the prediction and the ground truth could lead to an AVD of infinity. Therefore, we replaced the metric as $\min(1, AVD)$ to bound AVD. The complete calculation of the metrics was detailed in **Table 3**, and the evaluation was available at: https://github.com/hzlab/EPVS_challenge.

| **Metric** | **Formula** | **Range** | **Best Value** |
|---|---|---|---|
| DSC | $\dfrac{\lvert V_P \cap V_G \rvert}{\lvert V_p \rvert + \lvert V_G \rvert}$ | [0, 1] | 1 |
| AVD | $\min(1, \dfrac{\lvert V_p - V_G \rvert}{V_G})$ | [0, 1] | 0 |
| Recall | $\dfrac{I_P \cap I_G}{\lvert I_G \rvert}$ | [0, 1] | 1 |
| Precision | $\dfrac{I_P \cap I_G}{\lvert I_P \rvert}$ | [0, 1] | 1 |

**Notation:**
$V_p$ = Volume of the predicted segmentation
$V_G$ = Volume of the ground truth segmentation
$\lvert V_P \cap V_G \rvert$ = Intersection between $V_p$ and $V_G$
$\lvert V_p - V_G \rvert$ = Difference between $V_p$ and $V_G$
$\lvert I_p \rvert$ = Number of predicted instances
$\lvert I_G \rvert$ = Number of ground truth instances
$\lvert I_P \cap I_G \rvert$ = Number of correctly detected instances

**Table 3: Evaluation Metrics for 3D Segmentation Performance.** Abbreviations: DSC – dice similarity coefficient; AVD – absolute volume difference.

**2.5. Ranking**

In our evaluation framework, we assessed the performance of the methods in the validation and test phases with equal weighting, ranking each method based on four key metrics: DSC, AVD, Recall, and Precision.

For clarity, we notated $R_{i,j,v}$ the rank of the method $i$ in the metric $j$ in the validation phase, while $R_{i,j,t}$ denoted the same for the test phase. For each method $i$, we calculated its gross rank in the validation phase $G_{i,v}$ by averaging its ranks in the four metrics: $G_{i,v} = \frac{1}{4}\sum_{j=1}^{4} R_{i,j,v}$. Similarly, the gross rank in the test phase $G_{i,t}$ was computed as $G_{i,t} = \frac{1}{4}\sum_{j=1}^{4} R_{i,j,t}$. The final rank for each method $F_i$ was then determined by averaging its gross ranks from both phases: $F_i = \frac{G_{i,v} + G_{i,t}}{2}$.

This balanced evaluation approach ensured that methods demonstrate consistent performance in both validation and test phases while considering multiple performance aspects through the four metrics. The resulting final rank provided a comprehensive assessment of each method's relative standing within the competitive landscape, highlighting comparative strengths and weaknesses across all evaluation dimensions.

**2.6. Statistical Analysis**

All 4 metrics were calculated per scan and method in the validation and test phases. Pairwise comparisons of different methods were performed using the Wilcoxon paired tests for all 4 metrics in the ROIs of CSO and BG regions due to their non-normal distribution. For each method, the number of times it was found significantly better (with a p-value ≤ 0.05 for significance) than another was used to rank the given metric. The higher number of significant tests, the better the method.

**2.7. Participation and Submissions**

Ten teams submitted their methods during the validation phase, while seven teams submitted during the test phase. One of these seven teams failed to segment EPVS on most images of the validation and the test phases. Therefore, only six teams were considered for the awards. For both validation and test phase submissions, any missing results (i.e., cases where no output was generated) were assigned the worst possible metrics for that subject (e.g., a DSC of 0 and an AVD of 1). **Table 4** provides a comparative overview of the methodological approaches used by six participating teams, highlighting key differences in backbones, loss functions, and preprocessing steps. A summary of each method is provided below in alphabetical order.

- **bear_walker** bear_walker (Nanjing University, China) employed the nnUNet framework (Isensee et al., 2021) to train a 3D segmentation model for EPVS, achieving a DSC of 0.3189 on the validation set. The approach utilized a 3D U-Net architecture with leaky ReLUs and instance normalization instead of batch normalization. For preprocessing, images underwent statistical analysis on volume spacing and foreground intensity, with clipping at the 0.5 and 99.5 percentiles, followed by normalization and resampling to a target spacing of (0.94, 0.90, 0.94). The training employed a combination of Dice and cross-entropy losses with a 5-fold cross-validation strategy, using the Adam optimizer and an extensive set of data augmentation techniques, including random rotations, scaling, elastic deformations, and gamma correction.

- **BIG_AEHRC** The BIG_AEHRC team (CSIRO, Australia) implemented a modified MedNeXt model (Roy et al., 2023), adapted for both 3D and 2D segmentation of EPVS. Their approach intelligently handled varying slice thicknesses by applying 3D models to images with slice thicknesses smaller than 2.5mm and 2D models to thicker slices. MedNeXt, built on ConvNeXt blocks (Liu et al., 2022), leverages Residual Inverted Bottlenecks to maintain contextual information and employs the UpKern technique to iteratively increase kernel sizes, preventing performance saturation. The team trained models using 5-fold cross-validation with a Polynomial Learning Rate Scheduler and AdamW optimizer, applying Dice and TopK cross-entropy loss for 3D models and Dice and standard cross-entropy loss for 2D models. Their experiments also included comparisons with SwinUNETR (Hatamizadeh et al., 2022; Tang et al., 2022).

- **CCNU_CS_24** Team CCNU_CS_24 (Central China Normal University, China) introduced a U-Net-based multi-modal and multi-scale approach for EPVS segmentation. They implemented multi-branch coding by separately encoding single-modal data (T1w, T2w, T2-FLAIR) and incorporated multi-scale convolutional kernels (3×3, 5×5, 7×7) to enhance feature extraction for detecting the small, scattered EPVS structures. Their training process utilized SGD optimization with momentum of 0.9 and weight decay of 0.0001, combining Focal Loss and Dice Loss to address class imbalance issues inherent in EPVS segmentation. The team employed an ensemble prediction strategy combining three models: standard 3D U-Net, a model with separate modal encoding, and a model with multi-scale convolution kernels, using a pixel-by-pixel OR operation on individual predictions to generate the final segmentation.

- **HufsAIM** Team HufsAIM team (Hankuk University of Foreign Studies, Korea) implemented a complementary multi-view approach using four independent models: three 2D SwinUNETR (Hatamizadeh et al., 2022; Tang et al., 2022) models (axial, coronal, and sagittal views) and one 3D SwinUNETR model. This architecture employed Swin Transformer as the backbone, chosen specifically for its ability to capture both local and global features needed to detect the sparse, irregularly distributed EPVS structures. The models were trained with T1w, T2w, and T2-FLAIR inputs, using the AdamW optimizer (learning rate 0.0001, weight decay 0.001) and a combined Dice and cross-entropy loss function. For the final prediction, probability distributions from all four models were ensembled through averaging with equal weights, followed by applying an empirically determined threshold of 0.5. This approach demonstrated robustness to varying slice thicknesses (1-5mm) and effectively combined complementary information from different anatomical perspectives.
- **nic-vicorob** Team nic-vicorob (University of Girona, Spain) presented an ensemble approach based on the nnUNet framework (Isensee et al., 2021) adapted for EPVS segmentation. Their solution utilized 3D full resolution nnUNet models with ResidualEncoderUNet architecture enhanced with attention gates (Oktay et al., 2018) optimized under 5-fold cross-validation. Each model contained six levels with 3×3×3 convolution kernels and processes four input channels: T1w, T2w, T2-FLAIR, and a region of interest (ROI) mask combining BG and CSO regions. The team extracted 32 patches of size 64×64×64 per image, applied standard nnUNet preprocessing steps, including non-zero region cropping, z-score normalization, and voxel resampling (0.5 mm isotropic), and trained with SGD optimization using a combined Dice and cross-entropy loss function. They also explored synthetic data generation inspired by SynthSeg (Billot et al., 2023) though this approach was ultimately discarded due to overfitting issues.

- **SHIVA** The SHIVA team (Fealinx, France) presented SHIVA-PVS-v3, an enhanced model for EPVS segmentation based on their previously published approach (Boutinaud et al., 2021) but with important modifications. They implemented a custom 3D U-Net with residual blocks in TensorFlow, featuring 9 stages with two 3D convolutions at each stage, and replaced ReLU with Swish activation functions for smoother activations. Their model processed full brain volumes cropped to 160×214×176 voxels at approximately 1mm isotropic resolution. A key innovation was their dual-pathway approach that processes both T1w and intensity-inverted T2w images separately through five trained models from 5-fold cross-validation, then averages predictions by modality before taking the voxel-wise maximum between modalities. This strategy aimed to recover information potentially lost in registration, addressing challenges identified in previous research on PVS imaging (Ballerini et al., 2018; Duering et al., 2023; Zhu et al., 2010).

| Team | Backbone | Pretraining | Loss Functions | Input Modalities | Preprocessing | Augmentation | Post-processing |
|---|---|---|---|---|---|---|---|
| BIG_AEHRC | MedNeXt (2D & 3D) | No | Dice TopK CE (3D) CE (2D) | T1w T2w T2-FLAIR | Cropping | Unspecified | Restore to original space, threshold 0.3 (3D), 0.35 (2D) |
| nic-vicorob | 3D ResidualEncoderUNet with attention gates | No | Dice CE | T1w T2w T2-FLAIR ROI masks | Standard nnUNet preprocessing | Default nnUNet | Ensemble of 5 models, threshold 0.3 |
| HufsAIM | SwinUNETR (2D & 3D) | No | Dice CE | T1w T2w T2-FLAIR | Normalization | Random noise, random flipping | Ensemble of 2D & 3D models with threshold 0.5 |
| SHIVA | Custom 3D U-Net with residual blocks | Yes | Unspecified | T1w Inverted T2w | Cropping, resampling to 1 mm isotropic, intensity normalization | Random flipping, translation, rotation and intensity | Voxel-wise max of model averaging for two modalities, threshold 0.5 |
| CCNU_CS_24 | 3D U-Net with multi-scale convolution | No | Focal Loss Dice Loss | T1w T2w T2-FLAIR | Normalization | Image flipping | Ensemble of three models with OR operation |
| bear_walker | 3D nnUNet | No | Dice CE | T1w T2w T2-FLAIR | Default nnUNet | Random rotations, scaling, elastic deformation, gamma correction | Ensemble of five models |

**Table 4: Method Comparison for EPVS Segmentation. Abbreviation**: EPVS – enlarged perivascular spaces; CE – cross entropy loss.

Detailed information of each method can be found online at https://www.synapse.org/Synapse:syn64956514.

## 3. Results
### 3.1. Challenge Result

**Table 5** presents the final rankings of all participating teams. BIG_AEHRC emerged as the clear winner with the best overall rank (1.625), maintaining strong performance in both the validation (2.25) and the test phases (1.00). Notably, CCNU_CS_24 showed a significant improvement from validation (4.75) to test (2.75), resulting in the second-best final rank (3.750). Three teams (SHIVA, HufsAim, and nic-vicorob) had comparable final rankings around 4.00-4.63, though with different trajectories between the validation and the test phases. **Figure 4** presents a visual comparison of segmentation results from all top-performing methods across representative cases, highlighting characteristic error patterns in under-segmentation (false negatives) and over-segmentation (false positives).

The detailed validation phase performance metrics presented in **Table 6** reveal that BIG_AEHRC achieved the highest overall segmentation accuracy with DSC of 0.545 and 0.678 for CSO and BG regions, respectively. For the CSO region, HufsAim had the best recall (0.530), while thefridge demonstrated superior precision (0.804). In the BG region, BIG_AEHRC achieved the highest recall (0.697), and thefridge maintained the best precision (0.841). Notably, models of all teams showed better performance in BG regions compared to CSO, probably due to the more complex morphology of EPVS in the CSO.

| Team | Rank (Validation) | Rank (Test) | Final Rank |
|---|---|---|---|
| BIG_AEHRC | 2.25 | 1.00 | 1.625 |
| CCNU_CS_24 | 4.75 | 2.75 | 3.750 |
| SHIVA | 4.75 | 3.25 | 4.000 |
| HufsAim | 4.50 | 4.00 | 4.250 |
| nic-vicorob | 3.75 | 5.50 | 4.625 |
| bear_walker | 7.25 | 4.50 | 5.875 |
| GDVL | 10.00 | 7.00 | 8.500 |
| thefridge | 3.75 | | |
| AbdulQayyum | 6.25 | | |
| umb_zw | 7.75 | | |

**Table 5: Final rankings of teams in the EPVS segmentation challenge showing validation, test, and final rank scores.** Validation scores use green shades (darker green = better performance), and test scores use orange shades (darker orange = better performance).

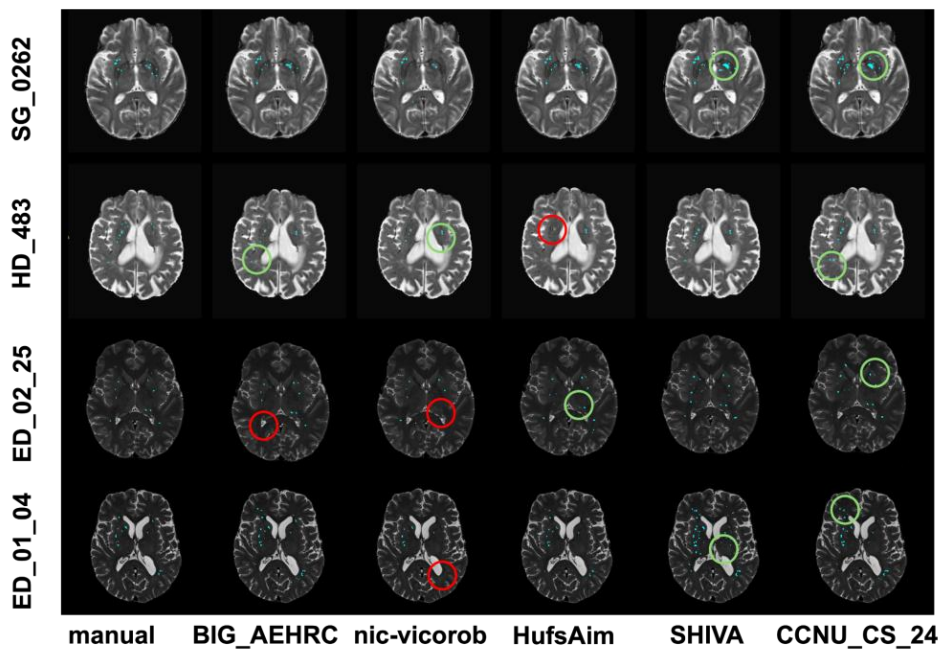

**Figure 4: Visual comparison of EPVS segmentation results of different methods on representative cases from each data site.** From left to right: manual segmentation (ground truth) and results from the top 5 performing methods BIG_AEHRC, nic-vicorob, HufsAim, SHIVA, and CCNU_CS_24. Red circles highlight under-segmentation (false negative), while green circles highlight over-segmentation (false positive). SG_0262 is from the SG70 cohort, HD_483 is from the MACC cohort while the other two are from the Edinburgh cohort. **Abbreviations:** EPVS – enlarged perivascular spaces.

| Team | CSO | | | | BG | | | |
|---|---|---|---|---|---|---|---|---|
| | Recall | Precision | DSC | AVD | Recall | Precision | DSC | AVD |
| BIG_AEHRC | 0.527 (2) | 0.713 (4) | 0.545 (1) | 0.235 (1) | 0.697 (1) | 0.750 (5) | 0.678 (1) | 0.243 (2) |
| thefridge | 0.316 (6) | 0.804 (1) | 0.429 (4) | 0.500 (5) | 0.527 (6) | 0.841 (1) | 0.631 (2) | 0.281 (5) |
| nic-vicorob | 0.385 (5) | 0.749 (2) | 0.469 (2) | 0.392 (4) | 0.527 (5) | 0.747 (6) | 0.621 (3) | 0.272 (4) |
| HufsAim | 0.530 (1) | 0.530 (9) | 0.457 (3) | 0.358 (2) | 0.627 (2) | 0.562 (9) | 0.579 (5) | 0.269 (3) |
| CCNU_CS_24 | 0.397 (4) | 0.629 (7) | 0.420 (5) | 0.374 (3) | 0.601 (4) | 0.639 (8) | 0.593 (4) | 0.221 (1) |
| SHIVA | 0.435 (3) | 0.676 (6) | 0.382 (6) | 0.507 (6) | 0.623 (3) | 0.820 (2) | 0.480 (7) | 0.713 (9) |
| AbdulQayyum | 0.207 (8) | 0.736 (3) | 0.269 (7) | 0.736 (8) | 0.397 (8) | 0.774 (3) | 0.486 (6) | 0.542 (7) |
| bear-walker | 0.219 (7) | 0.617 (8) | 0.222 (8) | 0.552 (7) | 0.428 (7) | 0.730 (7) | 0.397 (9) | 0.312 (6) |
| umb_zw | 0.186 (9) | 0.693 (5) | 0.219 (9) | 0.791 (9) | 0.361 (9) | 0.760 (4) | 0.421 (8) | 0.608 (8) |
| GDVL | 0.000 (10) | 0.001 (10) | 0.000 (10) | 0.998 (10) | 0.000 (10) | 0.000 (10) | 0.000 (10) | 0.999 (10) |

**Table 6: Performance metrics of different teams on the validation set of the EPVS segmentation challenge.** Each value is reported as Score (Rank). Background colors indicate ranking performance (darker green = better performance), with scores from only the top 5 ranking teams color-coded for each metric. **Abbreviations:** CSO – centrum semiovale; BG – basal ganglia; AVD – average volume difference; DSC – dice similarity coefficient.

**Figure 5** visualizes the performance distribution of all four metrics in the validation phase. The box plots reveal substantial variability in performance across teams, with BIG_AEHRC, nic-vicorob, and thefridge consistently ranking in the top positions. Several teams exhibited a wide distribution in their metrics, suggesting variable performance across different cases and anatomical regions.

In the test phase (**Table 7**), all teams experienced a performance drop compared to validation phase, reflecting the challenge of generalizing to unseen data. BIG_AEHRC maintained its leading position in all metrics despite reduced absolute scores (DSC of 0.326 for CSO and 0.441 for BG). CCNU_CS_24 showed remarkable robustness in the test phase, securing the second position in most metrics.

Team bear_walker demonstrated improved performance compared to validation, particularly in precision.

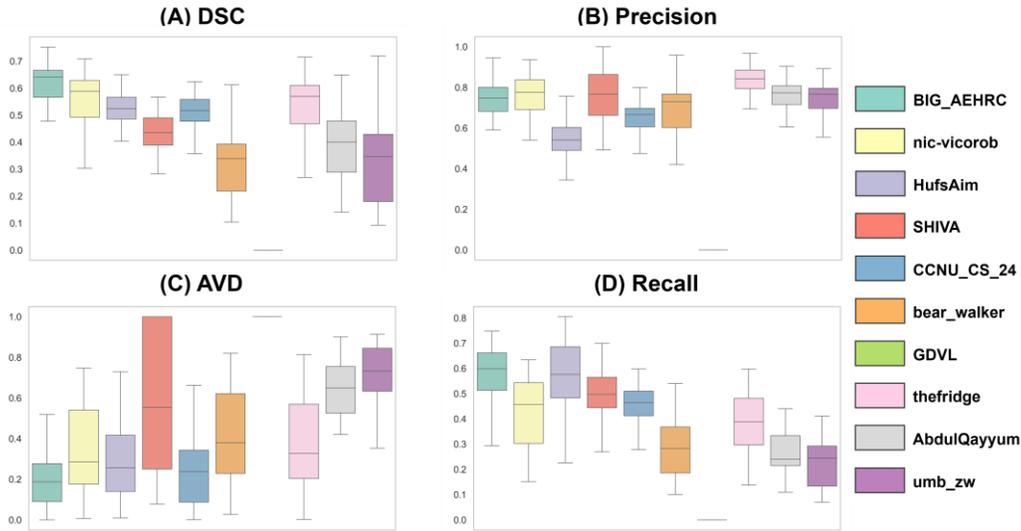

**Figure 5: Validation phase performance metrics across all teams for EPVS segmentation.** Box plots of DSC **(A)**, Precision **(B)**, AVD **(C)**, and Recall **(D)** are shown. Teams BIG_AEHRC, nic-vicorob and thefridge demonstrated the strongest overall performance, with BIG_AEHRC achieving the highest DSC (0.545 for CSO and 0.678 for BG regions). **Abbreviations:** AVD – average volume difference; DSC – dice similarity coefficient.

| Team | CSO | | | | BG | | | |
|---|---|---|---|---|---|---|---|---|
| | Recall | Precision | DSC | AVD | Recall | Precision | DSC | AVD |
| BIG_AEHRC | 0.316 (3) | 0.530 (1) | 0.326 (1) | 0.532 (1) | 0.514 (1) | 0.621 (1) | 0.441 (1) | 0.521 (1) |
| CCNU_CS_24 | 0.353 (1) | 0.365 (5) | 0.291 (2) | 0.587 (3) | 0.409 (3) | 0.494 (4) | 0.347 (2) | 0.606 (4) |
| SHIVA | 0.316 (4) | 0.405 (3) | 0.269 (3) | 0.648 (4) | 0.425 (2) | 0.511 (3) | 0.344 (3) | 0.685 (6) |
| HufsAim | 0.331 (2) | 0.354 (6) | 0.262 (4) | 0.543 (2) | 0.373 (4) | 0.441 (5) | 0.331 (4) | 0.589 (2) |
| bear-walker | 0.169 (5) | 0.412 (2) | 0.162 (6) | 0.697 (5) | 0.345 (5) | 0.512 (2) | 0.278 (6) | 0.596 (3) |
| nic-vicorob | 0.141 (6) | 0.388 (4) | 0.171 (5) | 0.793 (6) | 0.244 (6) | 0.410 (6) | 0.295 (5) | 0.661 (5) |
| GDVL | 0.000 (7) | 0.000 (7) | 0.000 (7) | 1.000 (7) | 0.000 (7) | 0.000 (7) | 0.000 (7) | 1.000 (7) |

**Table 7: Performance metrics of different teams on the test set of the EPVS segmentation challenge**. Each value is reported as Score (Rank). Background colors indicate ranking performance (darker orange = better performance), with scores from only the top 5 ranking teams color-coded for each metric. Abbreviations: CSO – centrum semiovale; BG – basal ganglia; AVD – average volume difference; DSC – dice similarity coefficient.

**Figure 6** illustrates the distribution of performance metrics in the test phase. The consistent superiority of BIG_AEHRC is evident in all metrics, with noticeably higher median values and reduced dispersion of results. The overall performance decline in the test phase is reflected in generally lower median values compared to the metrics in the validation phase in **Figure 5**.

For each method, the number of times it was found significantly better (with a p-value ≤ 0.05 for significance) than another in the validation phase and the test phase was shown in **Table 8** and **Table 9**, respectively, demonstrating strong alignment with the challenge rankings.

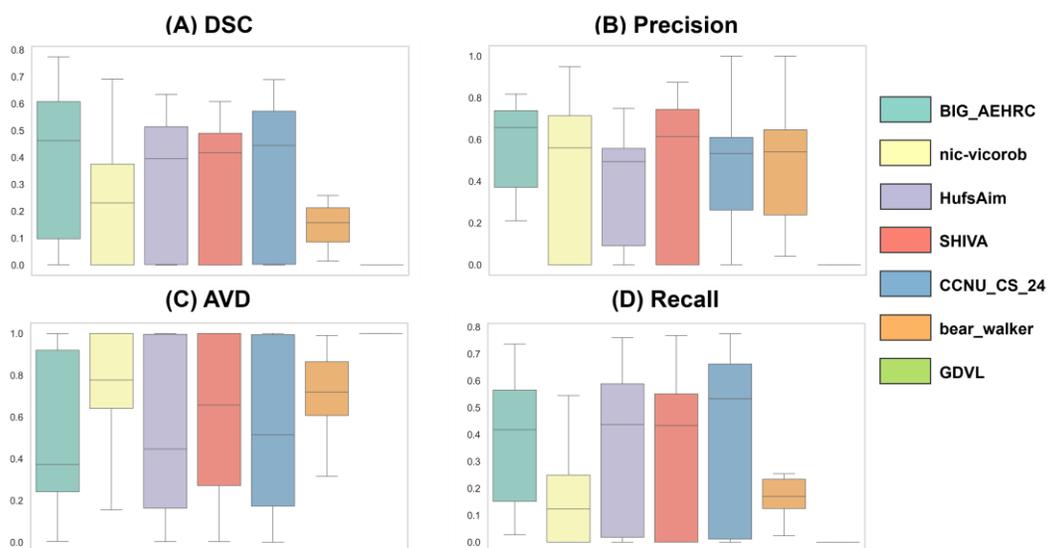

**Figure 6: Test phase performance metrics across the seven teams that completed the final evaluation.** Box plots of DSC **(A)**, Precision **(B)**, AVD **(C)**, and Recall **(D)** are shown. BIG_AEHRC maintained superior performance across all metrics, while other teams showed variable performance declines compared to the validation phase. **Abbreviations:** AVD – average volume difference; DSC – dice similarity coefficient.

| Team | CSO | | | | BG | | | | Total |
|---|---|---|---|---|---|---|---|---|---|
| | Recall | Precision | DSC | AVD | Recall | Precision | DSC | AVD | |
| BIG_AEHRC | 8 | 5 | 9 | 9 | 9 | 3 | 9 | 5 | 57 |
| thefridge | 4 | 9 | 5 | 4 | 4 | 8 | 8 | 5 | 47 |
| nic-vicorob | 5 | 7 | 7 | 5 | 4 | 3 | 7 | 5 | 43 |
| HufsAim | 8 | 1 | 6 | 6 | 7 | 1 | 5 | 4 | 38 |
| CCNU_CS_24 | 5 | 2 | 5 | 5 | 6 | 2 | 5 | 8 | 38 |
| SHIVA | 6 | 4 | 4 | 3 | 6 | 7 | 1 | 1 | 32 |
| AbdulQayyum | 2 | 5 | 3 | 2 | 2 | 4 | 3 | 3 | 24 |
| bear_walker | 2 | 2 | 1 | 3 | 3 | 3 | 1 | 4 | 19 |
| umb_zw | 1 | 4 | 1 | 1 | 1 | 3 | 2 | 1 | 14 |
| GDVL | 0 | 0 | 0 | 0 | 0 | 0 | 0 | 0 | 0 |

**Table 8: Statistical significance comparison of segmentation methods in the validation phase.** Each cell represents the number of times a team's method was found significantly better (p-value ≤ 0.05) than other methods for each metric and region. **Abbreviations:** CSO – centrum semiovale; BG – basal ganglia; AVD – average volume difference; DSC – dice similarity coefficient.

| Team | CSO | | | | BG | | | | Total |
|---|---|---|---|---|---|---|---|---|---|
| | Recall | Precision | DSC | AVD | Recall | Precision | DSC | AVD | |
| BIG_AEHRC | 3 | 6 | 6 | 4 | 5 | 6 | 6 | 6 | 42 |
| CCNU_CS_24 | 5 | 1 | 5 | 2 | 3 | 1 | 4 | 2 | 23 |
| SHIVA | 3 | 3 | 3 | 2 | 3 | 3 | 2 | 1 | 20 |
| HufsAim | 3 | 1 | 3 | 2 | 2 | 1 | 2 | 3 | 17 |
| bear_walker | 2 | 2 | 1 | 2 | 1 | 2 | 1 | 2 | 14 |
| nic-vicorob | 1 | 3 | 1 | 1 | 1 | 1 | 1 | 1 | 10 |
| GDVL | 0 | 0 | 0 | 0 | 0 | 0 | 0 | 0 | 0 |

**Table 9: Statistical significance comparison of segmentation methods in the test phase.** Each cell represents the number of times a team's method was found significantly better (p-value ≤ 0.05) than other methods for each metric and region. **Abbreviations:** CSO – centrum semiovale; BG – basal ganglia; AVD – average volume difference; DSC – dice similarity coefficient.

### 3.2. Additional Analysis

#### 3.2.1. Performance on different sites

To assess the generalizability of the methods, we analyzed the performance of the four datasets (**Figure 7**). Most algorithms showed consistent performance on the Edinburgh and Singapore sites (SG70 and MACC) that were represented in the training data. However, a significant performance drop was observed on the Shanghai TCM dataset, which was completely withheld during training and featured clinical acquisition parameters such as pixel and slice spacing. BIG_AEHRC demonstrated the most robust cross-site generalization, maintaining precision above 0.3 even on the Shanghai TCM data. CCNU_CS_24 and bear_walker also showed better adaptation to the new site compared to other methods. This analysis highlights the critical challenge of domain shift on developing EPVS segmentation algorithms that can be generalized across heterogeneous imaging protocols and scanner configurations. **Figure 8** illustrates a challenging case from the Shanghai TCM dataset where most methods failed to identify the EPVS structures, and only BIG_AEHRC detected partial EPVS structures. The high axial resolution and large spatial spacing of clinical scans from Shanghai TCM site may lead to different appearances of the EPVS structures, limiting the performance of existing segmentation algorithms trained on scans with different parameters.

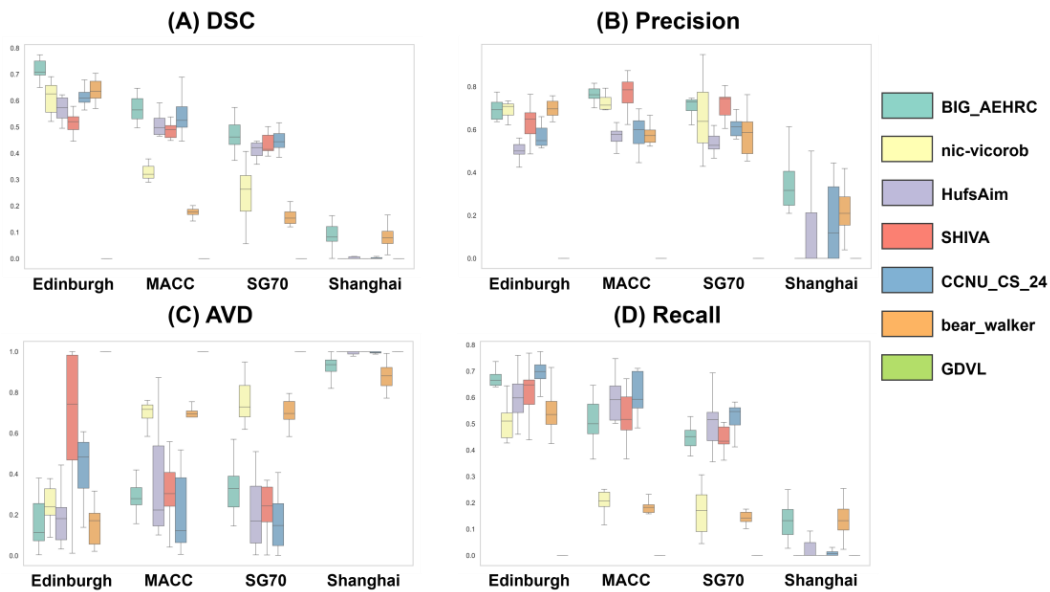

**Figure 7: Performance comparison across different sites (Edinburgh, NUS-SG70, NUS-MACC, and Shanghai TCM) for all metrics in the test phase.** Most algorithms showed decreased performance on the previously unseen Shanghai TCM dataset under a clinical setting, highlighting generalizability challenges across different acquisition sites and protocols. BIG_AEHRC maintained the most

consistent performance across sites. **Abbreviations:** AVD – average volume difference; DSC – dice similarity coefficient.

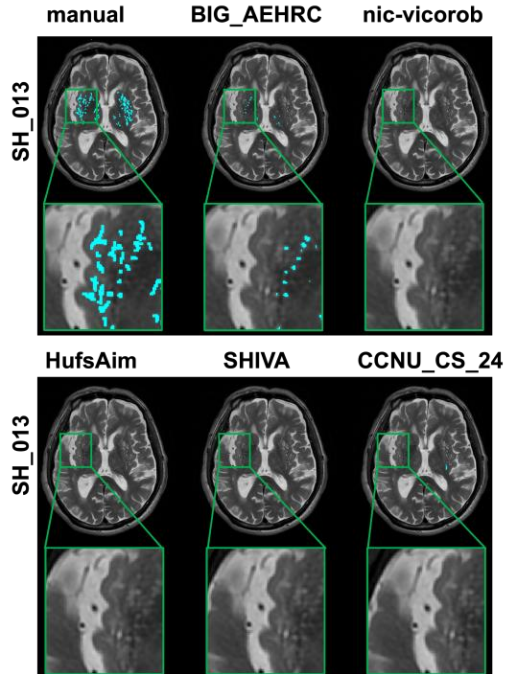

**Figure 8: Visualization of a challenging case from the Shanghai TCM dataset where most methods fail to identify the EPVS structures.** From left to right, top to bottom: manual segmentation (ground truth), and results from the top five performing methods. Only team BIG_AEHRC identifies partial EPVS structures. The green box highlights a zoomed-in region for better visualization of the fine EPVS structures.

### 3.2.2. Performance on different ROIs

The performance analysis by regions of interest (**Figure 9**) reveals that all teams achieved better segmentation in the BG compared to the CSO. The difference was particularly pronounced for the DSC, where the median improvement in BG over CSO ranged from 0.05 to 0.12 across teams. This disparity can be attributed to several factors: (1) EPVS in BG typically exhibits a more consistent morphology and orientation along the lenticulostriate arteries, (2) EPVS in CSO are more numerous, smaller and have more variable orientations, making them harder to delineate accurately, and (3) CSO regions often contain other small lesions of hyperintensities that can be confused with EPVS. The best-performing BIG_AEHRC method showed the smallest performance gap between regions, suggesting a more balanced feature extraction strategy across anatomical contexts.

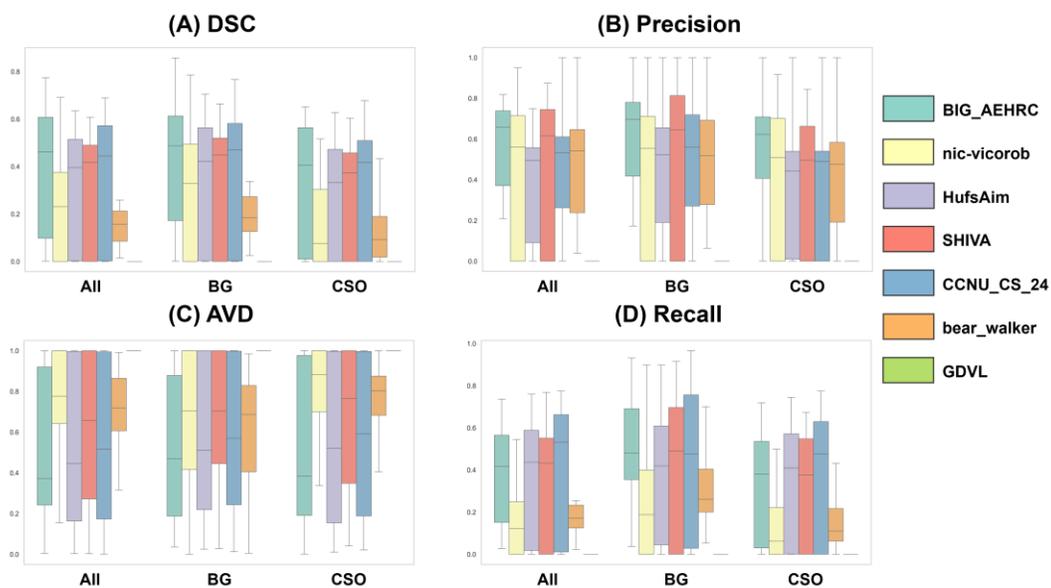

**Figure 9: Performance comparison across different brain regions (CSO and BG) for all metrics**. Most algorithms achieved better performance in BG regions compared to CSO, likely due to the more consistent morphology and higher contrast of EPVS in the BG. **Abbreviations:** CSO – centrum semiovale; BG – basal ganglia; AVD – average volume difference; DSC – dice similarity coefficient.

## 4. Discussion

EPVS segmentation is clinically impactful because EPVS burden is associated with CSVD, cognitive impairment, and other neurological conditions (Wardlaw et al., 2020; Paradise et al., 2021; Passiak et al., 2019). An accurate and automated solution could improve large-cohort analyses and early detection of disease-related changes without the burden of extensive manual annotations (Ballerini et al., 2018; Moses et al., 2023). Our EPVS challenge brought together multiple teams to develop automated methods for segmenting EPVS using multi-modal brain MRI. All methods submitted to this challenge were developed using deep learning methods, specifically variants of U-Net architectures, which aligns with current trends in other medical image segmentation applications as detailed in Section 2.3 and **Table 4**. Among all methods, the winning method BIG_AEHRC consistently outperformed others in both the validation and test phases, indicating robust performance and better cross-site generalizability as shown in **Table 5**. The top-performing methods demonstrated relatively good performance in terms of both individual EPVS identification measured by recall and precision and gross volume assessment measured by DSC and AVD when the test data was from the same cohort of the training dataset indicated by **Tables 6 and 7**, sharing similar scanner type, protocol, and demographics with the training data. BIG_AEHRC's approach, which combined MedNeXt architecture (Roy et al., 2023) with a dual 2D/3D strategy to handle varying slice thicknesses, proved particularly effective in maintaining consistent performance in both the validation and test phases for datasets like the training distribution. Team nic-vicorob relied on nnUNet (Isensee et al., 2021) but improved it with attention gates and an ROI mask; however, their performance dipped on the Shanghai TCM cohort due to the domain shift of voxel sizes and scanner settings. SHIVA's dual-pathway 3D U-Net by processing T1 and inverted T2 separately captured subtle EPVS features, but its reliance on robust registration could make it vulnerable to misalignment, especially when the voxel sizes of test data differ greatly from those from the training dataset. The HufsAim ensemble fused multiple views (axial, coronal, sagittal) plus a 3D Swin transformer (Hatamizadeh et al., 2022), improving recall at the cost of training complexity.

CCNU_CS_24 adopted a multi-scale approach and OR-based ensemble strategy to ensure high sensitivity, although it is prone to false positives. Meanwhile, bear_walker leveraged a straightforward 3D nnUNet pipeline with strong augmentations, but trailed behind top teams in overall segmentation quality. Lastly, GDVL encountered issues that resulted in failed segmentations in both validation and test phases, precluding deeper insights. Taken together, state-of-the-art backbones (e.g., MedNeXt) and carefully tuned ensemble approaches showed the best balance of generalizability, whereas other pipelines require additional domain adaptation to handle multi-site variability effectively. However, we observed a significant performance degradation when methods were applied to test sites where image parameters, such as pixel and slice spacing, differed substantially from the training set as illustrated in **Figure 7**. This was especially evident on the Shanghai TCM dataset which resembles the clinical setting, the performance of all methods severely degraded due to different appearances of EPVS structures on these scans (**Figure 8**). Even the best-performing method, BIG_AEHRC, while able to generate segmentation masks of moderate quality, showed marked decreases in all four metrics. These results demonstrate that current models are probably ready for research applications when the test data is of high isotropic resolution, but considerable challenges remain for cross-site generalizability and clinical applications where voxel sizes and scanner parameters differ significantly.

In addition, we consistently observed better performance in BG compared with the CSO in all methods as shown in **Figure 9**. This difference is likely due to the better contrast of EPVS in the BG, where these structures typically exhibit a more consistent morphology and orientation along the lenticulostriate arteries (Ballerini et al., 2018; Duering et al., 2023). In contrast, EPVS in CSO are sparse, smaller and have diverse appearances and variable orientations, making them harder to delineate accurately and differentiate from other white matter lesions as observed in **Figure 1**.

This challenge had several limitations that should be acknowledged. First, we worked with a relatively small and unevenly distributed dataset across different scanners and populations as detailed in Section 2.2 and **Table 2**, which restricted our capacity to thoroughly evaluate cross-site robustness. Future efforts should continue to work on curating high-quality annotated datasets as well as leveraging large-scale unannotated datasets in semi-supervised settings (Boutinaud et al., 2021) to improve model generalizability. Second, perivascular space labeling remains subjective and prone to variability, especially for small, sparsely distributed structures, creating a nontrivial source of ground-truth inconsistencies (as discussed in Section 2.3). Future work should focus on making multi-site annotation pipelines more consistent, including prescreening software, standardizing annotation tools such as the size of brush/pen, and facilitating frequent discussions on ambiguous annotation (Valdés Hernández et al., 2023).

In conclusion, this EPVS challenge represents a significant step forward for EPVS segmentation research. To the best of our knowledge, this challenge is pioneering in benchmarking diverse segmentation methods in a multi-site setting (Section 2.1). We identified some key winning approaches (such as combining state-of-the-art backbones with strategies for handling varying slice thicknesses) which could help further model development. At the same time, the challenge highlights the need for continued attention to developing robust cross-site generalization methods and standardizing annotation protocols to advance the field toward clinical applicability.

## 5. Funding

This study was supported by the Singapore Ministry of Health through the National Medical Research Council (NMRC) Office, MOH Holdings Pte Ltd under the NMRC Clinician Scientist Award (NMRC/CSASI/0007/2016), NMRC Clinician Scientist – Individual Research Grant (NMRC/CIRG/1485/2018 and MOH-001086), NMRC Singapore Translational Research Investigator Award (MOH-000707), NMRC Centre Grant (NMRC/CG/M009/2017_NUH/NUHS), Healthy Longevity Catalyst Awards (MOH-001441), NMRC Open Fund – Large Collaborative Grant (MOH-000500) and NMRC Open Fund – Individual Research Grant (MOH-001742), as well as the Research, Innovation and Enterprise (RIE) 2020 Advanced Manufacturing and Engineering (AME) Programmatic Fund from Agency for Science, Technology and Research (A*STAR), Singapore (No. A20G8b0102), Ministry of Education (MOE-T2EP40120-0007 & T2EP2-0223-0025, MOE-T2EP20220-0001), and Yong Loo Lin School of Medicine Research Core Funding, National University of Singapore, Singapore. Team NIC-VICOROB was supported by the Ministerio de Ciencia e Innovacion (PID2023-146187OB-I00) and by ICREA under the ICREA Academia programme. R.E.H. holds an IFUdG PhD Grant from University of Girona and V.A. hold an FPI grant (PRE2021-099121) from the Ministerio de Ciencia, Innovación y Universidades.

## 6. Declaration of competing interest

The author(s) declare that they have no known competing financial interests or personal relationships that could have appeared to influence the work reported in this paper.

## 7. Data availability

The EPVS Challenge Dataset is restricted to use within the EPVS Challenge 2024 only. Any additional use, including academic publications outside the challenge paper, requires formal data sharing agreement from the dataset owners. To request the challenge dataset, two separate Research Collaboration Agreements (RCA) with both National University of Singapore and the University of Edinburgh are required. Please contact co-organizers Dr Yichi Zhang (ychzhang@nus.edu.sg) and Dr Maria Valdés Hernández (m.valdes-hernan@ed.ac.uk) for the RCA details of both sites.

## Acknowledgments

We are particularly thankful to all the participants across all the datasets for their support of our research work. We also would like to thank the MICCAI 2024 challenge selection and organization committee for their strong support.

disease: associations with perivascular space volume and cognitive function. European Radiology, 34(2), 1314-1323. https://doi.org/10.1007/s00330-023-10122-3.

Zhu, Y.C., Dufouil, C., Soumaré, A., Mazoyer, B., Chabriat, H., Tzourio, C., 2010. High degree of dilated Virchow-Robin spaces on MRI is associated with increased risk of dementia. Journal of Alzheimer's Disease. 22(2), 663-672. https://doi.org/10.3233/JAD-2010-100378.

42